\documentclass[12pt]{article}
\input{amssym.def}
\input{amssym}
\usepackage[dvips]{color}
\usepackage{epsfig}
\usepackage{hyperref}
\normalbaselines
\makeatletter
\@addtoreset{equation}{section}
\makeatother

\def\be{\begin{equation}}
\def\ee{\end{equation}}
\def\ba{\begin{eqnarray}}
\def\ea{\end{eqnarray}}

\newcommand{\rf}[1]{(\ref{#1})}

\def\bra#1{\langle #1|}
\def\ket#1{|#1\rangle}

\def\green#1{\color{green}{#1}}

\begin{document}

\title{
Shared Information   in Stationary States at Criticality}
\author{Francisco C. Alcaraz$^1$  and Vladimir Rittenberg$^2$
\\[5mm] {\small\it
$^1$Instituto de F\'{\i}sica de S\~{a}o Carlos, Universidade de S\~{a}o Paulo, Caixa Postal 369, }\\
{\small\it 13560-590, S\~{a}o Carlos, SP, Brazil}\\
{\small\it$^{2}$Physikalisches Institut, Universit\"at Bonn,
  Nussallee 12, 53115 Bonn, Germany}
}
\date{\today}
\maketitle
\footnotetext[1]{\tt alcaraz@if.sc.usp.br}
\footnotetext[2]{\tt vladimir@th.physik.uni-bonn.de}

\begin{abstract}

We consider bipartitions of one-dimensional extended systems whose 
probability distribution functions 
describe stationary states of  stochastic models. We 
define estimators of the shared information between the two subsystems. If 
the correlation length is finite, the estimators stay finite for large 
system sizes. If the correlation length diverges, so do the estimators. 
The definition of the estimators is inspired by information theory. We 
look at several models and compare the behavior of the estimators in the 
finite-size scaling limit. Analytical and numerical methods as well as 
Monte Carlo simulations are used. We show how the finite-size scaling 
functions change for various phase transitions,  including the case where 
one has conformal invariance.

\end{abstract}

\newpage


\bigskip

\section{ Introduction} \label{sect1}

 In recent years a lot of attention has been paid to the understanding of the
entanglement properties of ground-state wavefunctions of hermitian
Hamiltonians describing one-dimensional quantum chain spin systems \cite{
CAR,CAR1}.
For a pure state if ${\cal C = A + B}$  is a bipartition, the von Neumann
entanglement entropy $S_{vN}$ is defined as:
\be \label{e1}
 S_{vN} ({\cal A}) = S_{vN} ({\cal B}) = 
-\mbox{Tr}(\rho_{\cal A}\ln\rho_{\cal A}),
\ee
where $\rho_{\cal A} = \mbox{Tr}_{\cal B}(\rho)$,   $\rho$ is the density matrix related to the 
ground-state wavefunction. If ${\cal C}$, ${\cal A}$ and ${\cal B}$ 
 have $L$, $l$ and respectively $L-l$
sites, for a large system and large subsystems, $S_q$ converges to a
constant if the system is not critical.  If the system is gapless and
conformal invariant, one gets
\be \label{e2}
  S_{vN}(l,L) \sim \gamma \ln l + C \quad(L>>l>>1),
\ee
and the finite-size scaling behavior
\be \label{e3}
 S_{vN} (l,L) = \gamma \ln \tilde{L}_C + C,\quad \tilde{L}_C= 
L\sin(\pi  l/L)/\pi ,
\ee
where $\gamma = c/6$ for an open system and $\gamma = 2c/6$ for periodic boundary conditions (c is the central charge of the Virasoro algebra). C is a
nonuniversal constant. The multipartite mutual information was also
studied \cite{CAR@}-\cite{CARA} and shown to give more information 
about the system beyond
the central charge.

 In the present paper we consider a bipartition of a large one-dimensional
interacting classical system ${\cal C}$ and study the shared information between
the subsystems ${\cal A}$ and ${\cal B}$ due to large scale correlations. 
Such systems can
be described by  probability distribution functions (PDF) of stationary states of stochastic models. They are ket ground-states
wavefunctions written in a special basis, of non hermitian and in general,
nonlocal quantum chain Hamiltonians which give the time evolution of Markov processes. The bra ground-states are trivial.
The essential difference between the classical and the quantum cases is
that in the former case the components of the ket wavefunctions are
non-negative real numbers which, when properly normalized, represent the
probabilities to find in the stationary state, various configurations of
the system.

 Wavefunctions with similar properties can also be encountered in the case of
some special hermitian Hamiltonians. An example is the ground-state 
  wavefunction of the spin 1/2 $SU(2)$ symmetric Hamiltonian if one uses the
valence bond ($SU(2)$ singlets) basis \cite{CTW,ACL}.

We present several estimators of shared information which have neat 
properties. They vanish if the subsystems are separated. 
Their values increase if the shared information increases. 
This happens if the systems are more constrained. 
It turns out that, in many cases, the estimators have a behavior 
which copies the one of the entanglement entropy. 
For example, if the PDF describes a system with a finite 
correlation length, the estimators stay finite for large systems. 
The area law \cite{ECP} is respected. 
If the system is critical one gets, in general,  logarithmic 
corrections to the area law. We are going to discuss in detail these 
corrections. (Some preliminary results were presented in \cite{ARS}.)

 In stationary states, the existence of divergent correlation lengths 
does not imply conformal invariance. 
Therefore one can have a larger variety of critical systems. 
We are going to show how the estimators behave in several cases. 
Without entering  into details here we just mention that if the 
evolution operator is conformal invariant, 
the estimators have the form 
\rf{e3} with different values for $\gamma$.

 Distinct estimators can be sensitive to distinct aspects of shared 
information. As in the case of quantum entanglement, 
in many cases the evaluation of estimators can be a very difficult task. 
For some estimators however, we can take advantage of the fact 
that we are dealing with stationary states of stochastic processes and use Monte Carlo simulations to compute them.

 The estimators of shared information are presented in Sec.~2. 
The configuration space in which the estimators are defined is given 
by Dyck paths. 
Dyck paths appear in different contexts. 
They are used in polymer modelling. They
 can also represent a one-dimensional interface. We also recall that 
if one takes a 
one-dimensional lattice, and on each lattice site one puts a 
spin 1/2 representation of $SU(2)$, 
to each $SU(2)$ singlet 
 corresponds  
 a Dyck path.
 This implies that the ground-state of a $SU(2)$ invariant 
one-dimensional quantum chain can be written in a Dyck paths basis.

 Different stochastic models have stationary states described by 
different PDF of the various Dyck paths.
 In Sec.~3 we present a simple local one-parameter stochastic model 
(see also Appendix A). The stationary states are given by a 
known model of polymer adsorption \cite{BEO}. 
Depending on the value of this parameter, one is in a gapped phase 
(finite correlation length) or in a gapless phase which is not 
conformal invariant. We compute the estimators in this model. In Appendix B, we compare the expression of the estimators with those 
obtained considering restricted and unrestricted Motzkin paths 
instead of Dyck paths.

 In Sec.~4 we consider a different non-local, 
one-parameter dependent, stochastic model: the raise and peel model \cite{GNPR,AR}. Depending on the value of the parameter, one is in a gapped phase, 
a conformal invariant phase or in a gapless, 
non-conformal invariant phase in which the critical exponents vary 
continuously with the parameter. 
Using analytic methods but mostly Monte Carlo simulations, 
we have computed the estimators and determined their expression in
 the finite-size scaling limit.

As pointed out in Refs.~\cite{XXB,XXC,XXA} in gapless conformal
invariant quantum chains, the R\'enyi entropies get different subleading
contributions if one consider separately systems with $l$ odd and $l$ even. In
Sec.~5 we discuss subleading contributions to several estimators in the
conformal invariant point of the raise and peel model.

 In Sec.~6 we change the configuration space from Dyck paths to ballot 
paths. The configuration space is not left-right symmetric anymore. 
We compute the finite-size scaling limit of two estimators 
in two different models and compare them with the left-right symmetric case.

 The estimators for the stationary states of the asymmetric exclusion 
process (ASEP) on a ring are computed in Sec.~7.

 We sum up our results and suggest other applications in Sec.~8.

\section{ 
 Estimators of shared information for Dyck paths configurations}

\begin{figure}[t]
\begin{center}
\begin{picture}(150,70)
\put(0,0){\epsfxsize=250pt\epsfbox{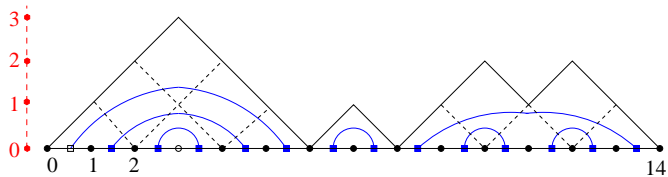}}
\end{picture}
\caption{
Example of a link pattern for $L = 14$ and the corresponding Dyck 
path. In the latter there are four contact points and three clusters. 
The shared information is the largest  in the left most cluster. 
The corresponding non-intersecting link pattern (blue lines) is also 
shown.}
\label{fig1}
\end{center}
\end{figure}

 We consider an open one-dimensional system with $L + 1$ 
sites ($L$ even). A
Dyck path (restricted solid-on-solid (RSOS) configuration) is defined as
follows. We attach to each site $i$ non-negative integer heights $h_i$
 which
 obey SOS rules:
\be \label{e4}
 h_{i+1} - h_i =\pm1, \quad h_0 = h_L = 0 \quad  (i = 0,1,\ldots,L-1).
\ee
There are
\be \label{e5}
Z_1 (L) = L!/(L/2)!(L/2 + 1)!
\ee
configurations of this kind. If $h_j = 0$ at the site $j$ one has a {\it 
contact point}.
 Between two consecutive contact points one has a {\it cluster}. There are
four contact points and three clusters in Fig.~\ref{fig1}. Contact points and
clusters play an important role in shared information.

 In the stochastic processes which we are going to define in the next two
sections, the Dyck path will be seen as a fluctuating interface between a
substrate ($h_{i} =(1 +(-1)^{i+1})/2$; $i = 0, \ldots,L$) 
 covered by tiles
(tilted squares) and a gas of tiles (not shown in Fig.~\ref{fig1}). 
The probability
distribution functions (PDF) of the various Dyck paths will be given by
the ground-state wavefunctions of Hamiltonians which give the continuum
time evolution of the systems.

 There is a known bijection between Dyck paths with $L + 1$ sites and
non-intersecting link patterns ($SU(2)$ singlets) for a one-dimensional spin 1/2
system with $L$ sites \cite{PMA}. 
In Fig.~\ref{fig1} one can see how this bijection is defined in 
a particular case. This 
observation opens the door for other possible applications of the 
present work. 
Spin 1/2 $SU(2)$ symmetric one-dimensional quantum chains have their 
dynamics ruled by Hermitian Hamitonians whose ground-state eigenfunctions 
are in the singlet representation. The XXX chain is an example.
If one studies the quantum 
 entanglement in this wavefunction, by taking the spin basis,    one 
looses the $SU(2)$ symmetry. On the other hand the $SU(2)$ singlet basis 
is not orthogonal. However it might happen that in the basis of $SU(2)$ 
singlets, the coefficients appearing in the wavefunction are non-negative 
real numbers which through a proper normalisation can be interpreted 
as probabilities \cite{CTW,ACL}. 
One can then consider the problem of shared information in the wavefunction as if it were the PDF of a stationary state of a stochastic process.

\begin{figure}[t]
\begin{center}
\begin{picture}(150,70)
\put(0,0){\epsfxsize=250pt\epsfbox{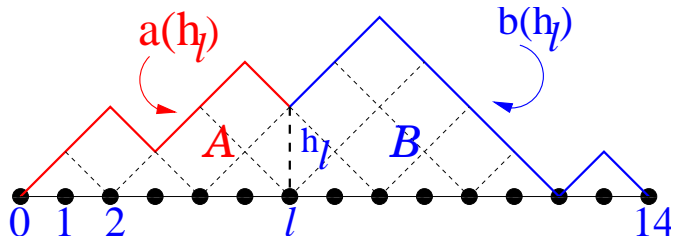}}
\end{picture}
\caption{
A Dyck path ($L=14$) with a height $h_l$ at the site $l$, splits into two 
ballot paths $a(h_l)$ and
$b(h_l)$ belonging to the subsystems $\cal A$ and $\cal B$.}
\label{fig2}
\end{center}
\end{figure}

 A bipartition of the system is obtained in the following way. The
ensemble of Dyck paths (system ${\cal C}$ with $L + 1$ sites) is divided into two
parts: the sites $0\leq i \leq l$ (part ${\cal A}$) and the sites 
$l\leq j \leq L$ (part ${\cal B}$).
This implies the splitting of each Dyck path, which at the site $l$ has the
height $h_l$, into two ballot paths. One RSOS path
$a(h_l)$ which starts at $i = 0$ with $h_i=0$  and ends at the site $l$ at the height $h_l$
 and another
one $b(h_l)$ which starts at $i = l$ with height $h_l$, and ends at  $i =L$ with $h_L=0$
 (see Fig.~\ref{fig2}).

 If we consider the Dyck paths which all have a contact point at $i = l$
($h_l = 0$), it is clear that for these paths, the subsystems ${\cal A}$ 
 and ${\cal B}$ are
separated. On the other hand, if we consider the Dyck paths which reach 
at $i = l$
a high value of $h_l$, they have around the site $l$ large clusters 
shared by
the two subsystems. This is due to the constraint given by the RSOS rules 
\rf{e4}.
The estimators we are going to define take into account these 
observations. They
all measure in different ways the amount of information that can be
obtained about the ballot paths in ${\cal B}$ if one observes ballot 
paths in ${\cal A}$.

 We denote by $P(a(h_l),b(h_l))$ the probability to have a given Dyck path
in ${\cal C}$ formed by the ballot paths $a(h_l)$ and $b(h_l)$ 
 in ${\cal A}$, respectively ${\cal B}$.
We consider the marginals
\be \label{e6}
 P(a(h_l)) = \sum_b P(a(h_l),b(h_l)),
\ee
and $P(b(h_l))$.
 The probability to have a height $h_l=h$ at the site $l$ is
\be \label{e7}
 P_l(h,L)  = \sum_a P(a(h_l)) = \sum_b P(b(h_l)).
\ee 

The proposed estimators are defined as follows:

{\it (I) Mutual Information,}

\be \label{e7m}
 I(l,L) = \sum_{h_l, a(h_l),b(h_l)} P(a(h_l),b(h_l)) \ln{\frac{P(a(h_l),b(h_l))}{P(a(h_l))P(b(h_l))}}.
\ee
This is the standard  estimator of shared information \cite{CTH} in 
information theory.

{\it (II) Interdependency}

This is the Shannon entropy in which the observable is the height at the
site $l$:
\be \label{e8}
H_h(l,L) = - \sum_{h} P_l(h,L)\ln P_l(h,L).
\ee
This estimator is new. It imitates the expression of the von Neumann entropy
 in which one uses the Schmidt decomposition. One can also define in a 
natural way the  R\'enyi interdependencies:
\be \label{e7p}
R_n (l,L) = 1/(1-n)\ln \sum_{h} P_l(h,L)^n, \quad n=2,3,\ldots.
\ee

{\it (III) Valence Bond Entanglement Entropy}

Is the average height at the site $l$:
\be \label{e9}
h(l,L) = \sum_{h} h P_l(h,L).
\ee
This estimator was introduced independently by Chhajiany et al \cite{CTW}
 and
Alet et al \cite{ACL} in the context of $SU(2)$ symmetric spin 1/2 quantum
chains, and further studied by Jacobsen and Saleur \cite{JLJ}. 
The aim was to
measure the average number of link ($SU(2)$ singlets) crossings 
at a given site 
of the quantum chain.

 We would like to stress that when computing the estimators, the cases of 
even (odd) values of $l$ have to be considered separately at least 
for small values of $l$ and $L$. The number of Dyck paths which have 
$h_l = 1$ ($l$ odd) for example, can be very different than the number 
of paths which have $h_l = 0$ ($l$ even). This phenomenon affects the 
values of \rf{e7}-\rf{e9} but becomes irrelevant for large values 
of $l$ and $L$.

 The estimators defined above are sensitive to the probabilities of 
having  
 large heights which implies, due to the RSOS rules, that one has 
large clusters and therefore large shared information.
The next two estimators, which are knew, 
 are sensitive to the probabilities to have 
zero heights. 

{\it {(IV)} Density of Contact Points Estimator}

\be \label{e10}
D(l,L) = - \ln P_l(0,L) \quad   (l \quad {\mbox{even}}).
\ee
 This estimator expresses the fact that if the density of contact points
 $\rho(l,L) = P_l(0,L)$ is small, one has, in general, large clusters.

{\it (V) Separation Shannon Entropy}

\be \label{e11}
 S(l,L) = H(L) - H(l) - H(L-l) ,    
\ee
where $H(M) = -\sum_k P_k\ln P_k$ is the Shannon entropy for a system of
size $M$ and $P_k$ is the probability to have a Dyck path $k$. 
The separation
 Shannon entropy measures the increase disorder in ${\cal C }$ 
as the effect of putting
together the subsystems ${\cal A}$ and ${\cal B}$.
All the estimators vanish if the subsystems $\cal A$ and $\cal B$ are 
separated. 

 It is easy to see that if all the Dyck paths have the same probabilities
equal to $1/Z_1(L)$,
\be \label{e12}
I(l,L) = H_h(l,L),
\ee
and
\be \label{e15}
D(l,L) = S(l,L) = \ln (Z_1(L)/Z_1(l)Z_1(L-l)).
\ee
To compute 
\rf{e12} we use 
\be \label{e13}
P_l(h,L) = d(l,h)d(L-l,h)/Z_1(L) 
\ee
in \rf{e8}, where  $d(l,h)$ is the number of ballot paths which start 
 from $h_0 = 0$ at $i=0$
and end at site $l$ with $h_l = h$.
 If $l$ is even, $h$ also takes even values. It is therefore convenient to
denote   $L = 2L'$, $l = 2l'$ and $h = 2 h'$. One has \cite{BEO}:
\be \label{e14}
d(l',h') =  \frac{2h'+1}{l'+h'+1}\frac{(2l)!}{(l'-h')! (l'+h')!}.
\ee
 The relations \rf{e12} and \rf{e15} stay valid for any model in which 
all configurations have the same probabilities (see also Section 6).

 In the next two sections we are going to compute and compare the
estimators of shared information in two very different models.

\section{ A stochastic model for polymer adsorption}

\begin{figure}[t]
\begin{center}
\begin{picture}(250,100)
\put(0,0){\epsfxsize=250pt\epsfbox{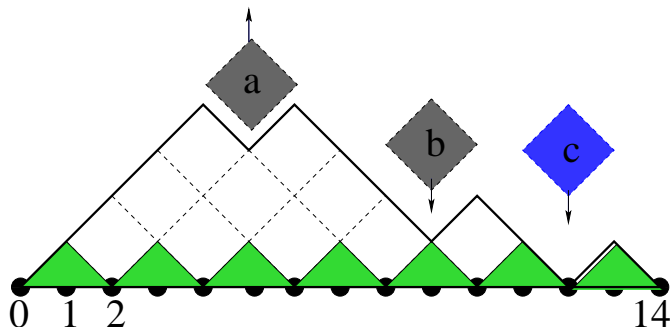}}
\end{picture}
\caption{
The stochastic model for 
polymer adsorption. The tilted square $a$ evaporates with a probability 
$q$. 
$b$ is adsorbed with a probability  $1$ while $c$ is adsorbed with a 
probability  $p$.}
\label{fig3}
\end{center}
\end{figure}

 We take as the configuration space for the stochastic model, the Dyck
paths with $L + 1$ sites, seen as an interface between a fluid composed
of tiles covering the substrate and a gas of tiles (Fig.~\ref{fig3}).
The model depends on a non-negative parameter $u$.

 Consider the slope variables $s_i = (h_{i+1} - h_{i-1})/2$. 
The evolution of
the system in discrete time (Monte Carlo steps) is given by the following
rules. With a probability $P_i = 1/(L - 1)$ each site $i$ of the system is
visited. If $s_i = \pm 1$ the interface (Dyck path) stays unchanged. 
If $s_i =0$ and $h_i > h _{i-1} > 0$ (local maximum, not part of the substrate), 
a tile
is removed ($h_i \rightarrow h_i - 2$) with a probability equal to $q$. If $s_i = 0$
 and $h_i <h_{i-1}$ (local minimum) a tile is added 
($h_i \rightarrow  h_i +2$) with a probability equal
to $p$ if $h_i = 0$ (the site $i$ is a contact point) and a 
 probabiblity to $q$
otherwise.  We take $p=u$, $q=1$ ($u<1$) and $p=1$, $q=1/u=K$ ($u>1$).

 For $u = 1$, this is the Rouse model \cite{ROS}, 
its extension for other values
of $u$ is so obvious that probably the model is known.
 It is a simple exercise to see that in the continuum time limit, the
Hamiltonian giving the time evolution of the system is non-hermitian except
for $u = 1$. 

The ground-state eigenfunction of the Hamiltonian has a very
simple expression: the components of all the configurations with the
same number of contact points are equal and are monomials in $u^{-1}$. If we take
the components of the configurations with no contact point equal to 1, the
components of the configurations with $m$ contact points are equal 
to $1/u^m$.
(In Appendix A one show the calculation in the case $L = 6$). The
normalization factor of the wavefunction is:
\be \label{e16}
Z_K(L) = \sum_{\psi} K^{m(\psi)},
\ee
where $K = 1/u$, $\psi$ are Dyck paths and $m(\psi)$ 
are the number of contact
points of the Dyck path. This is the partition function of the well-known
model of polymer adsorption \cite{AO} in which $K = \exp(\beta J)$
 is a Boltzmann
weight. Let us notice that if $u = K = 1$, all configurations have the same
probabilities and the Dyck path can be seen as a one-dimensional random
walker who starts at the origin and whose position is given by the height with  time running on
the horizontal axis.

 We denote by  $\rho_K(l,L)$  the density of contact points at the site $l$ for a
system with $L+1$ sites ($L$ even) 
and a Boltzmann weight $K$, and by 
$h_K(l,L)$
  the local
average height. Their averages over all the $L+1$ sites are:
\be \label{e17}
\bar{\rho}_K(L) = \frac{1}{L}\sum_l \rho_K(l,L), \quad \bar{h}_K(L)= \frac{1}{L} 
\sum_l h_K(l,L).
\ee
 Let us sum up some K-dependent properties of the model which can be 
obtained from asymptotic evaluations \cite{AO}.
\vspace{0.5cm}

\noindent{\bf $K < 2$}

\be \label{e18}
Z_K(L) \sim \frac {2K}{(2-K)^2} \sqrt{\frac{2}{\pi}} 2^L L^{-3/2},
\ee
\be \label{e19} 
\bar{\rho}_K(L) \sim \frac{2+K}{2-K} L^{-1},  \quad 
\bar{h}_K(L) \sim \sqrt{\frac{\pi}{8}} L^{1/2}.
\ee

\noindent{ \bf $K = 2$} 
\be \label{e21} 
Z_2 \sim \sqrt{\frac{\pi}{2}}2^L L^{-1/2}
\ee
\be \label{e22}
\bar{\rho}_2(L) \sim \sqrt{\frac{2}{\pi}} L^{-1/2}, \quad 
\bar{h}_2(L) \sim \sqrt{\frac{\pi}{32}}L^{1/2}.
\ee

\noindent {\bf $K > 2$}
\be \label{e24}
Z_K(L) \sim \frac{K-2}{K-1} \left(\frac {K}{\sqrt{K-1}}\right)^L,
\ee
\be \label{e25}
\bar{\rho}_K(L) \sim \frac{K-2}{2(K-1)}, \quad
\bar{h}_K(L) \sim \frac{K}{2(K-2)}.
\ee
 One notices that in the whole domain $0 < K < 2$ one is in the 
random walker
model ($K = 1$) universality class. Surprisingly, the average height 
has the same values for all $K$ in the whole region. Since there are few 
clusters, one expects the estimators of shared information to have 
large values for large system
 sizes.

 One has a phase transition at $K = 2$. At this value of $K$, 
the average value
of $\bar{h}_K(L)$ gets a value which is half the value observed for $K < 2$
 and the density of contact points decreases slower with the size of 
the system than for $0 < K <2$. 
Finally, for $K > 2$, the density of clusters is finite which implies
 that one has many small clusters and therefore one expects the 
estimators to saturate for large  system sizes.

 To illustrate how various estimators behave for large system sizes, we take
 the simplest example, taking $K = 1$. The probabilities of  the various 
Dyck paths are the same and equal to $1/Z_1 (L)$ (see \rf{e5}).
 We first compute the {\it mutual information} which in this case is equal to the {\it interdependency}  
(see \rf{e12}-\rf{e14}). 
It is convenient to take $L = 2L'$, $l = 2l'$ and $h = 2h'$.
Using the Stirling approximation one obtains
\be \label{e27}
P_l(h',L) \sim \frac{4}{\sqrt{\pi}}\frac{z^2 e^{-z^2}}{\sqrt{\tilde{L}_R}}, 
\ee
where  $z= h'/\sqrt{\tilde{L}_R}$ and 
\be \label{e27a}
\tilde{L}_R = l'(1-\frac{l'}{L'}) = \frac{\tilde{L}_{RW}}{2}.
\ee
Notice that $\tilde{L}_{RW}$ can be written as 
\be \label{e27b}
  \tilde{L}_{RW} = Lf(x), \quad f(x) = f(1-x), \quad f(x) \sim x \quad 
(x<<1),
\ee
with $x = l/L$ and $f(x)=x(1-x)$. We are going to prove that each universality class of 
critical behavior is characterized by a lenght $\tilde{L}$ with a 
distinct function 
$f(x)$.  
 From \rf{e8} and \rf{e27} we get
\be \label{e28}
I(l,L) = H_h(l,L) \sim \frac{1}{2} \ln\tilde{L}_{RW} +C,
\ee
where
\be \label{e29}
C = \gamma' - \frac{1}{2}\ln{\frac{\pi}{2}} - \frac{1}{2} \approx 
0.303007.
\ee
$\gamma'$ is the Euler constant.

The {\it R\'enyi interdependency} \rf{e7p} has the same asymptotic expression as the interdependency:
\be \label{e29p}
R_n(l,L) \sim  1/2 \ln \tilde{L}_R + C_n.
\ee
 From \rf{e27} we also get the expression of the {\it valence bond 
entanglement entropy} \rf{e9}:
\be \label{e30}
h(l,L) \sim \frac{4}{\sqrt{2\pi}} \tilde{L}_{RW}^{1/2}.
\ee
 Finally, one can compute the {\it separation Shannon entropy}, equal in this 
case to the {\it density of contact points} estimator, using \rf{e15} and the 
Stirling approximation 
\be \label{e31}
D(l,L)=S(l,L) \sim \frac{3}{2}\ln\tilde{L}_{RW} +\ln \frac{\pi}{2\sqrt{2}}.
\ee
The local density of contact points being:
\be \label{e32}
\rho_K(l,L) \sim \sqrt{\frac{8}{\pi}} \frac{1}{\tilde{L}_{RW}^{3/2}}.
\ee
 We observe that in the finite-size scaling regime all the estimators 
depend on the variable $\tilde{L}_R$, which for large values of $L'$ 
becomes 
equal to $l'$. This is similar to the finite-size scaling variable 
$L\sin(\pi l/L)/\pi$ appearing in section 5.

 We first notice that the valence bond entanglement entropy 
$h(l,L)$ behaves 
as a power for large distances. 
The other four estimators have the following behavior in the finite-size scaling limit:
\be \label{e32a}
E(l,L) \sim \gamma_E \ln \tilde{L}_{RW} + C_E,
\ee
where for $ 1<< l << L$,
\be \label{e32b}
E(l,L) \sim \gamma_E \ln l + C_E,
\ee
with $\gamma_I = \gamma_H  = 1/2$ and $\gamma_D = \gamma_S = 3/2$.

 An obvious question is: are the divergent part of the estimators of shared information universal?
This would imply that in the whole domain $0 < K < 2$, in the expressions 
\rf{e32a}-\rf{e32b} only the constants $C_E$ change. 
This was checked using Monte Carlo simulations for the interdepedency as 
well as for $R_2(l,L)$.  
In the case of the valence bond entanglement entropy $h(l,L)$, this implies that 
\rf{e30} is independent of $K$.

If one assumes for general $K$ the finite-size scaling form $h(l,L) = c_K 
\tilde{L}_{RW}^{1/2}$ than from the $K$-independence of $\bar{h}_K$ in 
\rf{e19}  we get that $c_K$ is independent of $K$, and hence universality. 
The same argument can be used in the case of $D(l,L)$
 using 
\rf{e32} and the $K$-independence of the power law fall off with the size of the system of $\bar{\rho}(K,L)$ given by \rf{e19}.

We show now that for large systems, the separation Shannon entropy is also universal.
 The Shannon entropy for a system of size $L$ is
\be \label{e33}
H_K = -\sum_{\psi} P_{\psi}\ln P_{\psi} = 
-\sum_m D(L,m) \frac{K^m}{Z_K(L)} \ln \left(\frac{K^m}{Z_K(L)}\right),
\ee
where $P_{\psi}$ is the probability of a Dyck path $\psi$, $D(L,m)$ 
is the number of Dyck paths with $m$ contact points, and
\be \label{e34}
Z_K(L) = \sum_mK^mD(L,m).
\ee
 From \rf{e33} and \rf{e34} one obtains:
\be \label{e35}
H_K(L) = \ln Z_K(L) - L \bar{\rho}_K(L) \ln K.
\ee
 We now use \rf{e18} and \rf{e19} to get the Shannon entropy for a 
system of size $L$
\be \label{e36}
H_K(L) = L\ln 2 -\frac{3}{2}\ln L - \frac{2+K}{2-K} \ln K 
+ \ln\left( \frac{2K}{(2-K)^2} \sqrt{\frac{2}{\pi}}\right).
\ee
 The 7eparation Shannon entropy \rf{e11} is:
\ba \label{e37}
S_K(l,L)&=& H_K(L) - H_K(l) - H_K(L-l) \nonumber \\
&=&\frac{3}{2}\ln \tilde{L}_{RW}+ \frac{2+K}{2-K} \ln K 
-\ln \left(\frac{2K}{(2-K)^2} \sqrt{\frac{2}{\pi}}\right),
\ea
where $\tilde{L}_{RW} = l(1-\frac{l}{L})$.
 We can conclude that the leading term is universal.

 There is a supplementary argument for the universality of the 
coefficient $\gamma_S = 3/2$. If instead of Dyck paths one takes, 
with equal probabilities, Motzkin paths (see Appendix B for their definition), the asymptotic expression of the partition function is:
\be \label{e38}
Z_M(L) = \sqrt{\frac{27}{4\pi}} \frac{3^L}{L^{3/2}},
\ee
with the same leading behavior as in \rf{e37}:
\be \label{e39}
S_K(l,L)\sim \frac{3}{2}\ln \tilde{L}_{RW} + \frac{1}{2}\ln\left(\frac{4\pi}{27}\right).
\ee
 This is to be expected  since one has again a random walker who, 
at each step, can not only move up or down but can also stay in place. 
This corresponds, in the continuum,  only to a change of the diffusion 
constant.
 More interestingly, in Appendix B we consider restricted Motzkin paths. 
They correspond to $SU(3)$ link patterns \cite{FZJ}. Because of the 
constraint one  expects more shared information and a larger 
value for $\gamma_S$. 
This is indeed the case. One finds $\gamma_S = 4$.

 We did not compute the mutual information for values of $K\neq 1$ but we assume that one  gets $\gamma_I = 1/2$ for the whole interval $0 < K < 2$.

 To sum up, in the whole critical domain all five estimators of shared 
information have a universal, albeit different, large lattice 
sizes behavior.
We did  not study properly the $K = 2$ case. Heuristic arguments suggest 
that the shared information is smaller than in the case of the 
random walker 
universality class ($K < 2$). Comparing $h_2(L)$ 
 given by \rf{e22} with $h_K (L)$ given by \rf{e19}, for $K < 2$,  
  we conclude that the valence bond entanglement 
entropy for $K = 2$ is half of the $K < 2$ value. From the expression 
\rf{e22} of the average density of contact points one expects a value 
$\gamma_D = 1/2$ as compared with $\gamma_D = 3/2$ for $K < 2$.

 We discuss now the $K > 2$ domain in which the correlation length is 
finite.
In this domain one has a finite density of small clusters 
and a finite averaged height (see 
 \rf{e25}). 
 This implies that all the estimators have a finite value for large 
lattice sizes. This can be seen directly for the valence bond entanglement 
entropy and for  
 the density of contact points estimator 
by using  \rf{e25}, and the separation Shannon entropy 
by using  \rf{e11}, \rf{e24}, \rf{e25} and \rf{e35}. 
The finiteness of the remaining two estimators comes from the 
exponential fall off of the probabilities for large size systems.

\section{
 Shared information in the Raise and Peel model}

This model has been studied in detail in several papers (\cite{GNPR,APR}, for a review see\cite{AR}). We are going to mention here only those facts which are relevant for the study of the shared information.

 Like in the previous section, we consider the interface  separating a film of tiles deposited on the substrate, from a rarefied gas of tiles. The model depends on a non-negative parameter $u$ and the evolution of the system in discrete time  (Monte Carlo steps) is given by the following rules. With a probability $P_i = 1/(L-1)$ a tile from the  gas hits the site $i =1,\ldots,L-1$.

Depending of the slope $s_i=(h_{i+1}-h_{i-1})/2$ 
at the site $i$, the following processes can occur:

\noindent 1) $s_i = 0$ and $h_i > h_{i-1}$. The tile hits a peak and is reflected.

\noindent 2) $s_i = 0$ and $h_i < h_{i-1}$. The tile hits a local minimum and is  adsorbed ($h_i\rightarrow  h_i + 2$) with a probability $p$ and is reflected with a probability $1 - p$.

\begin{figure}[t]
\begin{center}
\begin{picture}(250,100)
\put(0,0){\epsfxsize=250pt\epsfbox{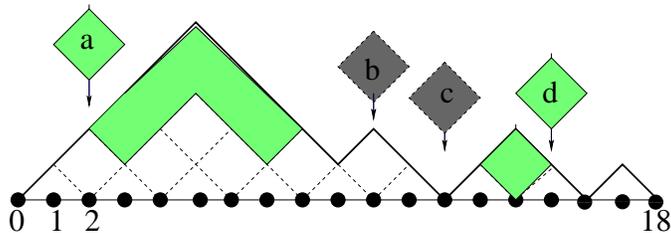}}
\end{picture}
\caption{
Example of a configuration of the interface in the raise peel model for 
 $L = 18$. Depending on the position where the tilted tiles reach the 
interface,  several distinct processes occur (see the text).}
\label{fig4}
\end{center}
\end{figure}

\noindent 3) $s_i = 1$. The tile is reflected after triggering, with probability $q$, the desorption ($h_j \rightarrow h_j-2$) of a layer of $b-1$ tiles from the segment $\{j=i+1,\ldots,i+b-1\}$ where $h_j>h_i=h_{i+b}$
 (see Fig.~\ref{fig4}). With a probability $1-q$ nothing happens.

\noindent  4) $s_i = -1$. With probability $q$, the tile is reflected after 
triggering the desorption ($h_j \rightarrow h_j-2$) of a layer of 
$b-1$ tiles belonging to the segment $\{j=i-b+1,\ldots,i-1\}$ 
where $h_j>h_i=h_{i-b}$. With a probability $1-q$ nothing happens.

We take $p=u$, $q=1$ for $u\leq 1$ and $p=1$, $q=1/u$ for $u\geq1$. 
In Fig.~4 we illustrate the possible processes in the case of a Dyck 
path with $L=18$. The tilted tile $b$ corresponds to process 1), tile $c$ 
to process 2), tile $a$ to process 3) and tile $d$ to process 4).

\begin{figure}[t]
\begin{center}
\begin{picture}(260,120)
\put(0,0){\epsfxsize=250pt\epsfbox{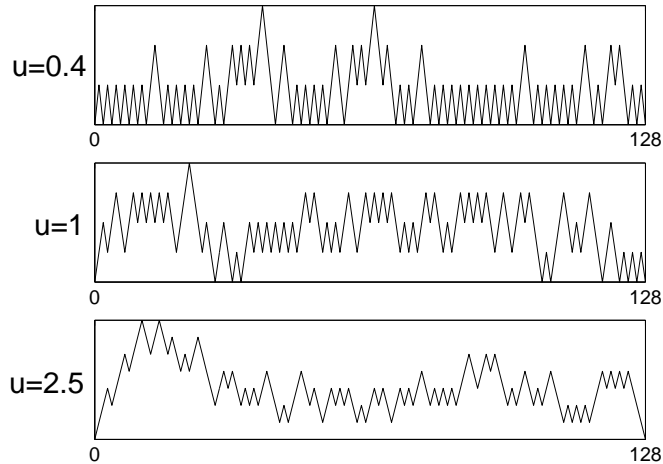}}
\end{picture}
\caption{
  Typical configurations in the stationary states for
$ L+1= 129$ sites and three  values of $u$. }
\label{fig5}
\end{center}
\end{figure}

 The phase diagram of the model was determined from the study of the finite-size scaling limit eigenspectrum of the Hamiltonian which gives the time evolution of the system, and other methods. For $0 < u < 1$, the system is massive. For $u = 1$, the system is conformal invariant with an effective central charge of the Virasoro  algebra $c = 0$. 
For this case only, the stochastic process is described by a spin 1/2 XXZ quantum chain with $U_q(sl(2))$ symmetry with $q=\exp(i\pi/3)$. The link 
pattern corresponds to $U_q(sl(2))$ singlets \cite{PMA}. 
For $u > 1$, the system is scale invariant but not conformal invariant. The dynamic critical exponent $z$ decreases continuously from the value $z = 1$ when $u = 1$ to very small values when 
$u$ becomes large. 
 What makes this model special is the fact that it contains at $u = 1$ a stochastic model which is conformal invariant and we can therefore  compare the behavior of the estimators with those studied in the problem of quantum entanglement.
 The dynamical properties of the model are reflected in the properties of the stationary state. For $0 < u < 1$, the densities of clusters are finite for large system sizes (like in the polymer adsorption model for $u < 1/2$). For all values of  $u \geq 1$ the density of clusters vanishes with a power which varies continuously with $u$. This was not the case in the polymer adsorption model where for all values of $u > 1/2$, one is  in the random walker phase. Some typical configurations for three values of $u$ are shown in 
Fig.~\ref{fig5} for a lattice of $L = 128$ (129 sites). One notices many clusters for $u = 0.4$ (in the massive phase), fewer clusters for $u = 1$ and a single cluster for $u = 2.5$. One finds fewer and fewer clusters when $u$ increases.

 We discuss the shared information in the stationary states of the model. In the domain $0 < u < 1$, the density of clusters is finite for large values of $L$ and so are the average values of the heights. As discussed in the previous section, this implies that all the estimators stay finite for large $L$. The situation is different when the system is critical. 

 We first discuss the $u = 1$ case.
 We will show that all the estimators of shared information have the following
expression in the finite-size scaling limit:
\be \label{e51}
E(l,L) \sim \gamma_E \ln\tilde{L}_C + C_E,
\ee
with
\be \label{e52}
\tilde{L}_C = L\sin(\pi l/L)/\pi, 
\ee
from which follows
\be \label{e53}
E(l,L) \sim \gamma_E\ln l + C_E, \quad   1<< l << L.
\ee
These expressions are similar to what was found for the von Neumann
entanglement entropy 
\rf{e2}.

 The {\it mutual information} 
\rf{e7m} was obtained using the exact PDF of the
stationary states up to $L = 26$ and assuming the finite-size scaling behavior \rf{e51}.
 The data are compatible with \rf{e51} if $\gamma_I = 0.07$ and
$C_I = 0.65$. This is a useful estimate rather than a result.

 In order to proceed, using Monte-Carlo simulations, we have determined the
probabilities $P_l(h,L)$ of having a height $h$ at the site $l$ for a lattice of size
$L$ and stored the information. This is possible since even for $L = 40000$
  the probability to obtain a height larger than 70 is negligible. These data
have been used for several purposes.

\begin{figure}[t]
\begin{center}
\begin{picture}(240,150)
\put(0,0){\epsfxsize=250pt\epsfbox{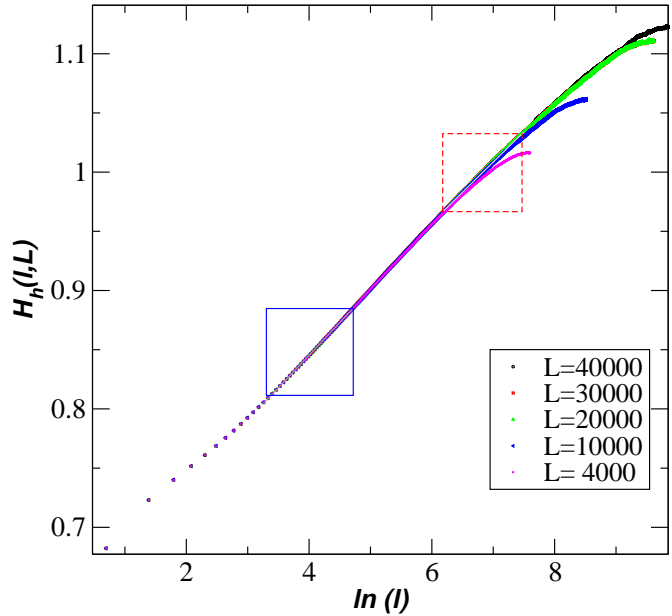}}
\end{picture}
\caption{
 The interdependency $H_h(l,L)$ as a function of $\ln l$ for various
lattice sizes $L = 4000,10000, 20000,30000, 40000$. 
Also shown  the two
domains where two linear fits were made in order to determine 
$\gamma_H$ and $C_H$.}
\label{fig6n}
\end{center}
\end{figure}

\begin{figure}[t]
\begin{center}
\begin{picture}(270,250)
\put(0,0){\epsfxsize=250pt\epsfbox{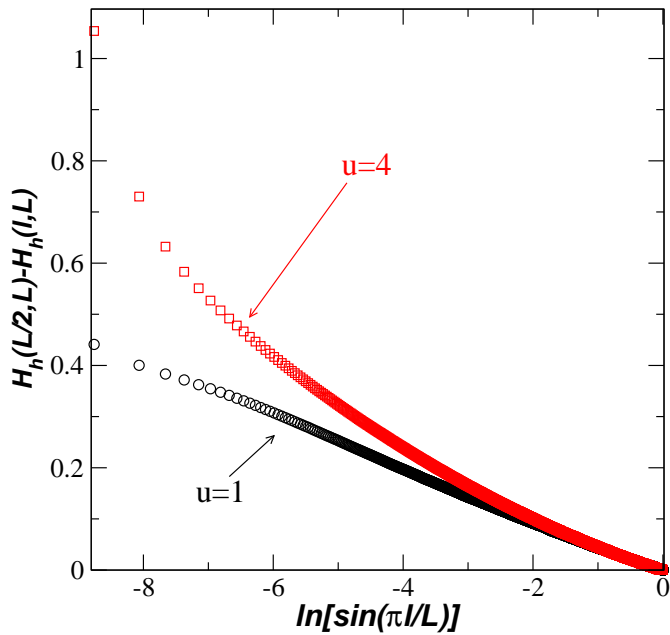}}
\end{picture}
\caption{
 $H_h(L/2,L) - H_h(l,L)$ as a function of $\ln(\sin(\pi l/L))$ for 
$u = 1$ and
$u = 4$ ($L = 40000$).}
\label{fig6}
\end{center}
\end{figure}

 We first studied the {\it interdependency} $H_h(l,L)$. 
To determine $\gamma_H$ we have
taken large lattices (up to $L = 40000$) and $1<<l << L$. In Fig.~\ref{fig6n} we show the plot
$H_h(l,L)$ as a function of $\ln l$ for various lattice sizes. 
As expected we have
data collapse. The larger the values of $L$, 
the larger is the domain of data
collapse. We have fitted the $L = 40000$ data assuming straight lines 
in two
regions shown by the two squares of Fig.~\ref{fig6n}. 
One obtains $H_h = 0.055 \ln l + 0.62$
for the low values of $\ln l$ and $H_h(l,L) = 0.051 \ln l + 0.65$ 
for the larger
 $\ln l$ domain. 
We take the  result of the second fit for
our estimates of $\gamma_H$ and $C_H$ since probably, 
in the first domain the
condition $1 << l$ was not satisfied.
 Maybe not surprisingly, the values for $\gamma_H$ and $C_H$ obtained 
for the
interdependency are close to those obtained for the mutual information 
$\gamma_I$
and $C_I$. As shown in Sec.~3, if the probabilities for all 
configurations are
equal (this is not the case here), the mutual information and the
interdependency are equal.
 In order to find the expression of the interdependency in the 
finite-scaling
limit, we have taken a large lattice ($L = 40000$) and computed the 
quantity
($H_h(L/2,L) - H_h(l,L)$) as a function of $\ln(\sin(\pi l/L))$. 
If the finite-size
scaling expression 
\rf{e51} is correct, one should obtain a straight line with a
slope equal to $\gamma_H$. In Fig.~\ref{fig6} we show the results of the 
Monte Carlo
simulation and one can see that in the finite-size scaling limit, $H_h(l/L)$
 has
the expression \rf{e51} for $u=1$, which is not the case for  
 $u=4$.

 The {\it R\'enyi interdependency} $R_n(l,L)$ were also measured. We have found
$\gamma_{R_2}  = 0.05$, $C_{R_2} = 0.4$ and $\gamma_{R_3} = 0.04$, 
$C_{R_3} = 0.4$. 
We have
to stress that a precise determination of the values of $\gamma_{R_n}$
 for the R\'enyi
interdependencies using Monte Carlo simulations is not an easy task. The
values of $l$, for which one can assume that the finite-size 
scaling limit is valid, are large. 
This can be seen from the
non-negligible differences between the R\'enyi interdependencies computed 
for $l$
odd and $l$ even
 due to subleading terms in 
\rf{e51}. In Sec.~5 we
discuss in detail these subleading contributions.

 The {\it valence bond entanglement entropy} $h(l,L)$ was also measured. We have found $\gamma_h = 0.277$ and $C_h = 0.75$. 
The determination of $\gamma_h$ and the check of the finite-size  
scaling function
\rf{e51} are much more precise in the case of $h(l,L)$ as shown in Figure 
2 of
\cite{ARS}.  Non-leading effects play a smaller role.
Considering periodic boundary conditions and the boundary Coulomb gas formalism, Jacobsen and Saleur \cite{JLJ} have computed $\gamma_h$ in a related model. Their value divided by 2 (we have an open system while in their 
case the chain is periodic) is $\gamma = \sqrt{3}/2\pi \approx  0.275$, very close to the measured value $\gamma_h$. 
This observation is not trivial because it proves that valence bond 
entanglement entropy is a {\it bona fide} estimator. 
In the periodic system 
the information is shared at the two ends of the subsystems.

 Moreover, we have determined the second   cumulant $\kappa_2$  of 
the 
probability distribution       $P_l(h,L)$, it has the same 
finite-size scaling behavior as the estimators:
\be \label{donkey}
\kappa_2(l,L) \sim \beta_2 \ln\tilde{L}_C +  b_2. 
\ee
We found $\kappa_2 =0.19$ and $b_2 = 0.25$  is in excellent 
agreement with half the value
   obtained by Jacobsen and Saleur \cite{JLJ}: $(2\pi\sqrt{3} - 9)/\pi^2 \approx 0.190767$. The fact that $\kappa_2$ in our case (open system) is 
half the values obtained in \cite{JLJ}, which corresponds to a periodic 
system, indicates that $\kappa_2$ can also be an estimator for the shared 
information between  subsystems.  

 The {\it density of contact points} estimator was computed using an almost rigorously derived expression 
 \cite{APR} for the density of contact points: 
\be \label{e41}
\rho(l,L) \sim \frac{\alpha}{\tilde{L}_C^{1/3}}
\ee
where
\be \label{e42}
\alpha = - \frac{\sqrt{3}\Gamma(-\frac{1}{6})}{6\pi^{5/6}}.
\ee
Using 
\rf{e10} we have:
\be \label{e43}
D(l,L) \sim -\ln \rho(l,L) = \frac{1}{3} \ln \tilde{L}_C  + 0.28349.
\ee

 The {\it separation Shannon entropy} was estimated  using the same data as for  the mutual information. We found $\gamma_S = 0.4$ and 
$C_S = 0.7$. 

 We notice that the values of $\gamma_D$ for the density of contact points estimator and $\gamma_S$ for the separation Shannon entropy are close to each other but very different of those determined for the mutual information and the interdependency.
One has to keep in mind that $D(l,L)$ and $S(l,L)$ coincide if the 
configurations
have the same probabilities.

 To sum up, in the conformal invariant point of the model, all estimators have the expression 
\rf{e51} with several values for $\gamma$ (see Table~\ref{table1} ). 
\begin{table}[hbt]
\centering
\begin{tabular}{|c||c|c||c|c|}
\hline
 &     $u=1$    &  $u=1$  &  $u=4$  & $u=4$ \\
 &    $\gamma_E$   & $C_E$  &    $\gamma_E$  & $C_E$    
  \\ \hline 
  Mutual information &    0.07  &    0.65 &  - & -  \\ \hline
  Interdependency    &    0.050   &   0.67   &   0.09 & 0.91 \\ \hline
  R\'enyi ($n=2$)    &    0.05   &   0.39   &  0.06 & 0.09 \\ \hline
  Valence bond ent.  &  0.277   &  0.75   &   0.63 & 1.37 \\ \hline
  Dens. Contact points & 0.333  &  0.284 &  0.73 & 0.71 \\ \hline
  Separation Shannon &    0.4    &   0.7 & - & - \\ \hline
\end{tabular}
\caption[table1]{
The values of $\gamma_E$ and $C_E$ in 
\rf{e51} and \rf{e52}  }
\label{table1}
\end{table}

 We have studied the estimators of shared information for several values of $u > 1$. We are presenting the results for $u = 4$ and just mention what happens at other values of $u$.
 For $u = 4$ the system stays critical, the dynamic critical exponent 
is $z \approx 0.3 $ and  conformal invariance is lost ($z \neq 1$). 

\begin{figure}[t]
\begin{center}
\begin{picture}(200,250)
\put(0,0){\epsfxsize=250pt\epsfbox{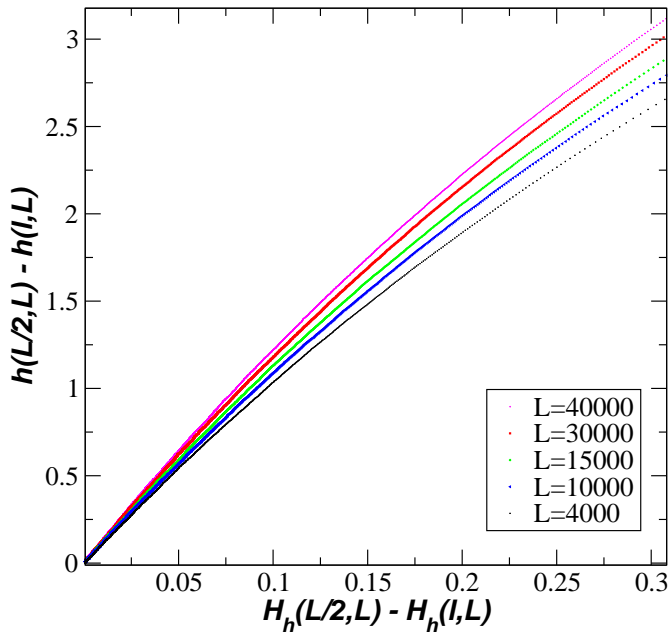}}
\end{picture}
\caption{
 ($h(L/2,L) - h(l,L)$) as a function of ($H_h(L/2,L) - H_h(l,L)$). 
The
 data were obtained
from lattices sizes $L = 4000, 10000, 15000, 30000$ and $40000$.}
\label{fig8n}
\end{center}
\end{figure}

 From the Monte Carlo simulations, the following picture has emerged. 
Equation   
\rf{e53} stays valid but in the finite-size scaling limit one has, 
instead of \rf{e52}, 
\be \label{e54}
E (l,L) \sim \gamma_E \ln \tilde{L}_{u=4}   + C_E,
\ee
where the $\tilde{L}_{u=4}=Lf_{u=4}(l/L)$ function has replaced the function $\tilde{L}_C = \tilde{L}_{u=1}$. 
That one has a new
finite-size scaling function can be seen, for 
example,  from Fig.~\ref{fig6}  
and from Figure 2  of
\cite{ARS}. 
The fact that there is only one finite-size scaling function
for all the estimators can be seen, for example,  from Fig.~\ref{fig8n}. 
Taking different values of $l/L$, we have plotted
the differences between the values of two estimators 
($H(l,L)$ and $h(l,L)$) at
$l = L/2$ and $l$. 
If there is only one scaling function one should obtain a
straight line with a slope given by the ratio $\gamma_h/\gamma_H$. 
One can see
that for large values of $L$, this is the case.

 The values of $\gamma_E$ obtained for $u = 4$ are given in Table 1. 
We notice that
they are all larger  when compared with the 
values
at $u = 1$. The shared information is larger. 
We notice that the values of $C_E$
 also increase.

We have explored also different values of $u$. 
The larger the value of $u$, the
larger are the values of $\gamma_E$ and $C_E$. 
The shape of the finite-size scaling
function $\tilde{L}_u$ changes continuously with $u$.


%
\section{
 Parity effects and subleading contributions to estimators in the raise 
and peel model at the conformal invariant point.}

As noticed in Sec.~4, for small lattices there are important differences 
between the R\'enyi interdependencies computed for $l$ odd and $l$ even at 
$u = 1$.
A similar phenomenon was observed for the R\'enyi entropies but not for 
the von Neumann entanglement entropy, in the study 
of quantum chains \cite{XXB,XXC,XXA}. Moreover, in \cite{XXC} it was 
conjectured that 
for the XXZ model with an anisotropy $-1\leq \Delta \leq 1$ and open 
boundaries, in the 
finite-size scaling limit, one has:
\be \label{X.1}
\delta R_n(l,L) = f_n(l/L)/\tilde{L}_c^{K/n}
\ee
where 
\be \label{X.2}
\delta R_n(l,L) = R_n(l,L) - R_n(l+1,L) \quad  (l \, \, {\mbox{even}})  
\ee
\be \label{X.3}
R_n(l,L) = \frac{1}{1 - n} \ln  {\mbox{Tr}} (\rho_A^n)
\ee
and
\be \label{X.4}
K = \pi/(2\arccos(\Delta)).  
\ee
$\rho_A$ is defined in 
\rf{e1} and $\tilde{L}_c$ in \rf{e51}. In \rf{X.1} $f_n(0)$ is finite. 
This
conjecture was checked analytically at the decoupling point ($\Delta = 0$) and
numerically for other anisotropies. For periodic boundary conditions, $K$ 
has to be replaced by $2K$ in \rf{X.1}.

 We have decided to look to the odd-even finite-size corrections for some 
of the estimators discussed in Sec~4 and see if one gets a behavior 
similar to \rf{X.1}. We have considered the quantities
\be \label{X.5}
\delta E(l,L) = E(l,L) - E(l+1,L) \quad  (l\,\, {\mbox{ even}})   
\ee
where $E(l,L)$ is an estimator.

\begin{figure}[ht]
\begin{center}
\begin{picture}(200,250)
\put(0,0){\epsfxsize=250pt\epsfbox{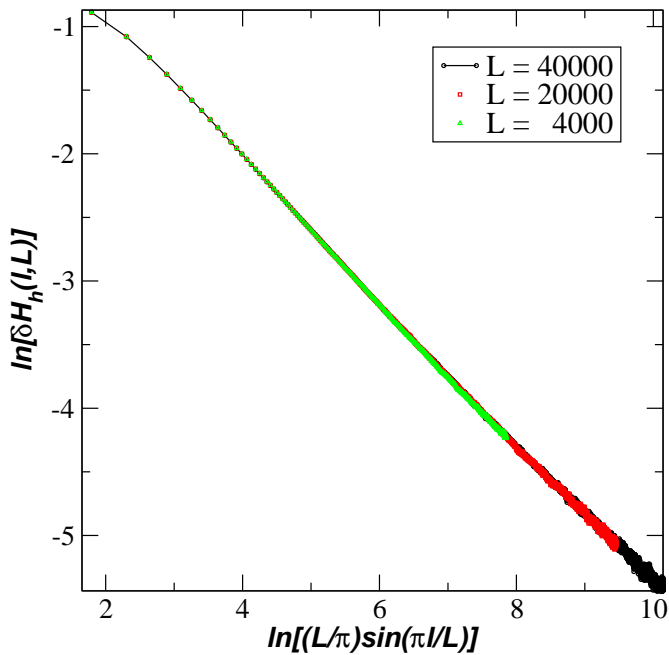}}
\end{picture}
\caption{
  Odd-even corrections to the interdependency (see \rf{X.5} and 
\rf{X.6}. We have used lattices of sizes $L = 4000, 20000$ and $40000$.}
\label{X1}
\end{center}
\end{figure}

 In the case of the interdependency, the data shown in Fig.~\ref{X1} suggest 
an  
expression similar to \rf{X.1}:
\be \label{X.6}
\delta H \sim  c_1/\tilde{L}_c^{x_1}        
\ee
with $c_1 \approx  1.3$ and $x_1 \approx  0.58$. The value of the exponent $x_1$,
 as 
expected,  does not match the exponent $K = 3/2$ ($n = 1$, $\Delta = 1/2$ \cite{GNPR} ) in
\rf{X.1}. The bases in the quantum chain and in the raise and peel model are 
completely different and there is no reason to have any connection between 
the exponents.   

\begin{figure}[ht]
\begin{center}
\begin{picture}(200,250)
\put(0,0){\epsfxsize=250pt\epsfbox{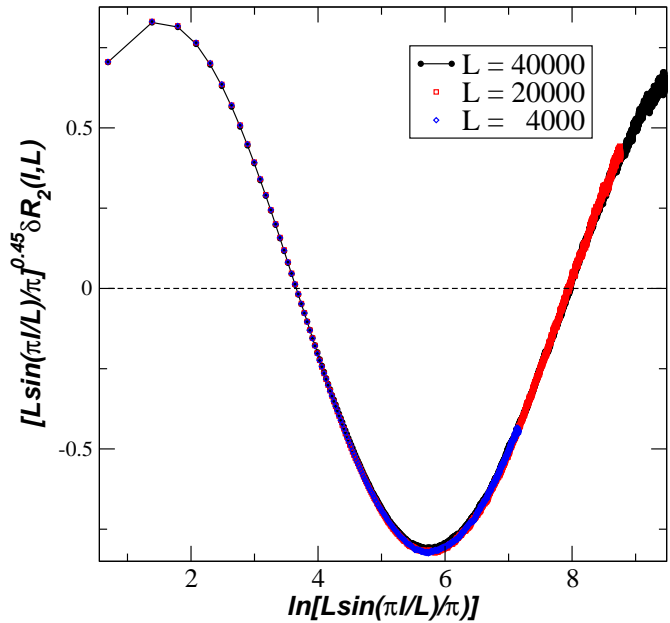}}
\end{picture}
\caption{
The odd-even corrections for the $R_2(l,L)$ R\'enyi
interdependency multiplied by $\tilde{L}_c^{0.45}$, a a function of $\ln \tilde{L}_c$  for different lattice sizes ($L =
4000$, $20000$ and $40000$).}
\label{X2}
\end{center}
\end{figure}

\begin{figure}[ht]
\begin{center}
\begin{picture}(200,250)
\put(0,0){\epsfxsize=250pt\epsfbox{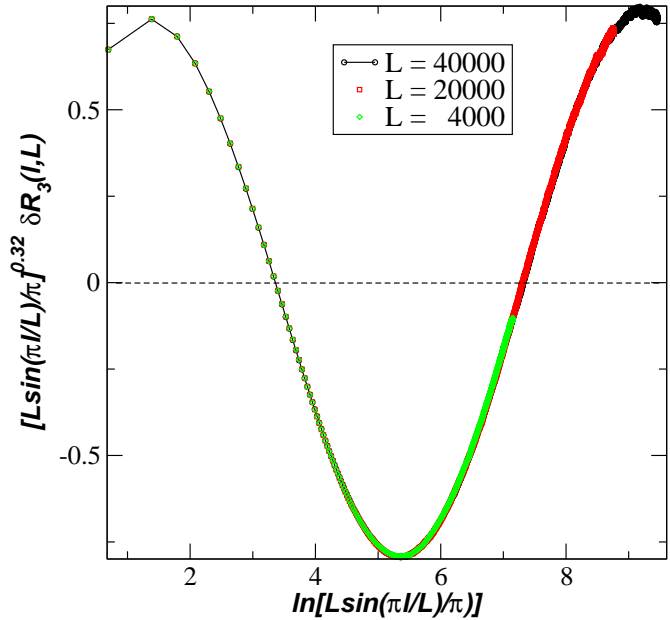}}
\end{picture}
\caption{
The odd-even corrections for the $R_3(l,L)$ R\'enyi
interdependency multiplied by $\tilde{L}_c^{0.32}$, a a function of $\ln \tilde{L}_c$  for different lattice sizes ($L =
4000$, $20000$ and $40000$).}
\label{X3}
\end{center}
\end{figure}

 A new phenomenon, not seen is the quantum chains, appears if one 
considers the R\'enyi interdependencies \rf{e7p}. We have measured 
$\delta R_2(l,L)$  
and $\delta R_3(l,L)$ using lattices up to $40000$ sites. The data shown in 
Figs.~\ref{X2} and \ref{X3} suggest the following expressions for the odd-even 
corrections:
\be \label{X.7}
\delta R_n \sim  1/{\tilde L}_c^{x_n} F_n(\ln(L_c)) 
\ee
Where $x_2\approx 0.45$, $x_3 \approx 0.32$  and $F_n(y)$ are oscillating functions. This 
observation came as a surprise. It is important to stress that this is not 
an effect of the finite-size scaling limit ($l/L$ finite). Oscillations are 
seen also in the limit $L$ very large and $l$ finite. 
Notice that like in \rf{X.1} we have $x_1>x_2>x_3$. 

 We have no explanation for this oscillatory behavior 
and it would be interesting 
to know if this is a pathology of the raise and peel model at the 
conformal point or the phenomenon is more general. We have looked at the 
odd-even corrections at higher values of $u$. The corrections become small 
and it is hard to make statements based on Monte Carlo simulations.
\begin{figure}[ht]
\begin{center}
\begin{picture}(200,250)
\put(0,0){\epsfxsize=250pt\epsfbox{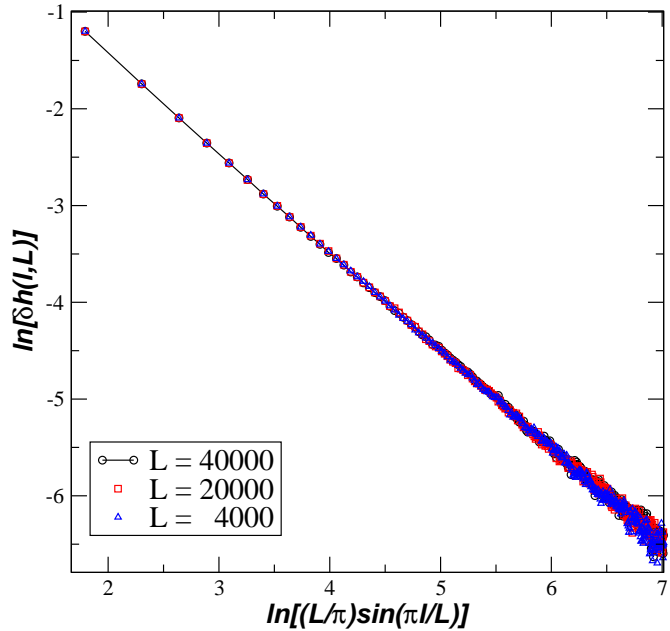}}
\end{picture}
\caption{
 Odd-even corrections to the valence bond entanglement entropy. 
Lattices of sizes  $L = 4000, 20000$ and $40000$ were used.}
\label{X4}
\end{center}
\end{figure}

 We have also looked at the odd-even corrections for the valence bond 
entanglement entropy \rf{e9} and found (see Fig.~\ref{X4})
\be \label{X.8}
\delta h(l,L) = c_h/\tilde{L}_c^{x_h} 
\ee 
with $c_h \approx 0.50$ and $x_h \approx 0.99$. No oscillatorry behavior
 was seen.

Odd-even effects are also seen for the estimators of shared
information in the case of the polymer adsorption model but we are not
going to discuss them here.

\section{
 Shared information in non left-right symmetric stationary states.}

The two stochastic models considered in the last sections were left-right symmetric. As a consequence, the expressions of the estimators in the finite-size scaling limit were of the form:
\be \label{e1.1}
E(l,L) \sim \gamma_E \ln(Lf(x)) + C_E
\ee
where $f(x) = f(1 - x)$ and $f(x) \approx x $, for small values of $x$. In this Section we consider a stochastic process with a source at the left-end of the system in order to see how the asymmetry in the system is reflected in the estimators of shared information. We will consider two examples only.   
 
 Instead Dyck paths, we consider ballot paths. Those are paths respecting the RSOS rules \rf{e4} with a modification: $h_L$ is fixed like for 
Dyck paths ($h_L = 0$) but $h_0$ is free ($h_0 = 0, 2,\ldots,L$). See 
Fig.~9 for an example. There are $L!/((L/2)!)^2$ configurations 
of this kind. 
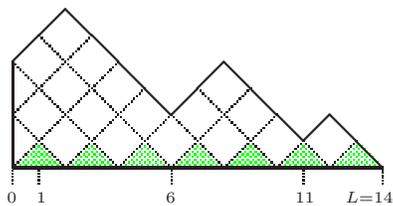
\begin{figure}[t] \label{fig7}
\begin{center}
\begin{picture}(250,70)(0,-10)
{\green{
\qbezier[8](13,11)(17,15)(21,19)
\qbezier[7](15,11)(18.5,14.5)(22,18)
\qbezier[6](17,11)(20,14)(23,17)
\qbezier[5](19,11)(21.5,13.5)(24,16)
\qbezier[4](21,11)(23,13)(25,15)
\qbezier[3](23,11)(24.5,12.5)(26,14)
\qbezier[2](25,11)(26,12)(27,13)
\qbezier[1](27,11)(27.5,11.5)(28,12)
\qbezier[8](33,11)(37,15)(41,19)
\qbezier[7](35,11)(38.5,14.5)(42,18)
\qbezier[6](37,11)(40,14)(43,17)
\qbezier[5](39,11)(41.5,13.5)(44,16)
\qbezier[4](41,11)(43,13)(45,15)
\qbezier[3](43,11)(44.5,12.5)(46,14)
\qbezier[2](45,11)(46,12)(47,13)
\qbezier[1](47,11)(47.5,11.5)(48,12)
\qbezier[8](53,11)(57,15)(61,19)
\qbezier[7](55,11)(58.5,14.5)(62,18)
\qbezier[6](57,11)(60,14)(63,17)
\qbezier[5](59,11)(61.5,13.5)(64,16)
\qbezier[4](61,11)(63,13)(65,15)
\qbezier[3](63,11)(64.5,12.5)(66,14)
\qbezier[2](65,11)(66,12)(67,13)
\qbezier[1](67,11)(67.5,11.5)(68,12)
\qbezier[8](73,11)(77,15)(81,19)
\qbezier[7](75,11)(78.5,14.5)(82,18)
\qbezier[6](77,11)(80,14)(83,17)
\qbezier[5](79,11)(81.5,13.5)(84,16)
\qbezier[4](81,11)(83,13)(85,15)
\qbezier[3](83,11)(84.5,12.5)(86,14)
\qbezier[2](85,11)(86,12)(87,13)
\qbezier[1](87,11)(87.5,11.5)(88,12)
\qbezier[8](93,11)(97,15)(101,19)
\qbezier[7](95,11)(98.5,14.5)(102,18)
\qbezier[6](97,11)(100,14)(103,17)
\qbezier[5](99,11)(101.5,13.5)(104,16)
\qbezier[4](101,11)(103,13)(105,15)
\qbezier[3](103,11)(104.5,12.5)(106,14)
\qbezier[2](105,11)(106,12)(107,13)
\qbezier[1](107,11)(107.5,11.5)(108,12)
\qbezier[8](113,11)(117,15)(121,19)
\qbezier[7](115,11)(118.5,14.5)(122,18)
\qbezier[6](117,11)(120,14)(123,17)
\qbezier[5](119,11)(121.5,13.5)(124,16)
\qbezier[4](121,11)(123,13)(125,15)
\qbezier[3](123,11)(124.5,12.5)(126,14)
\qbezier[2](125,11)(126,12)(127,13)
\qbezier[1](127,11)(127.5,11.5)(128,12)
\qbezier[8](133,11)(137,15)(141,19)
\qbezier[7](135,11)(138.5,14.5)(142,18)
\qbezier[6](137,11)(140,14)(143,17)
\qbezier[5](139,11)(141.5,13.5)(144,16)
\qbezier[4](141,11)(143,13)(145,15)
\qbezier[3](143,11)(144.5,12.5)(146,14)
\qbezier[2](145,11)(146,12)(147,13)
\qbezier[1](147,11)(147.5,11.5)(148,12)
}}
\qbezier[30](10,30)(25,45)(40,60)
\qbezier[40](10,10)(30,30)(50,50)
\qbezier[30](30,10)(45,25)(60,40)
\qbezier[20](50,10)(60,20)(70,30)
\qbezier[30](70,10)(85,25)(100,40)
\qbezier[20](90,10)(100,20)(110,30)
\qbezier[10](110,10)(115,15)(120,20)
\qbezier[10](130,10)(135,15)(140,20)
\qbezier[20](10,30)(20,20)(30,10)
\qbezier[40](10,50)(30,30)(50,10)
\qbezier[50](20,60)(45,35)(70,10)
\qbezier[20](70,30)(80,20)(90,10)
\qbezier[30](80,40)(95,25)(110,10)
\qbezier[10](120,20)(125,15)(130,10)
{\thicklines
\put(10,10){\line(1,0){140}}
\put(10,10){\line(0,1){40}}
\put(10,50){\line(1,1){20}}
\put(30,70){\line(1,-1){40}}
\put(70,30){\line(1,1){20}}
\put(90,50){\line(1,-1){30}}
\put(120,20){\line(1,1){10}}
\put(130,30){\line(1,-1){20}}
}
\qbezier[5](10,4)(10,7)(10,10)
\qbezier[5](20,4)(20,7)(20,10)
\qbezier[5](70,4)(70,7)(70,10)
\qbezier[5](120,4)(120,7)(120,10)
\qbezier[6](150,4)(150,7)(150,10)
\put(8,-3){$\scriptscriptstyle 0$}
\put(19,-3){$\scriptscriptstyle 1$}
\put(68,-3){$\scriptscriptstyle 6$}
\put(117,-3){$\scriptscriptstyle 11$}
\put(136,-3){$\scriptscriptstyle L=14$}
\end{picture}
\parbox{15cm}{
\caption{\small  A ballot path for $L = 14$. One has one contact point and one
cluster of size 14.}
}
\end{center}
\end{figure}

 Assume that each of these configurations has equal probability. One can easily compute the finite-size scaling limit of the {\it density of contact 
points} estimator (equal to the separation Shannon entropy), one obtains:
\ba \label{e1.2}
&& D(l,L) = S(l,L) 
\sim \frac{3}{2} \ln\left[l(1-\frac{l}{L})(\frac{l}{L})^{1/3}\right] + \frac{1}{2}\ln 
\left(\frac {\pi}{8}\right) \nonumber \\
&&=\frac{3}{2}\ln \tilde{L}_{RW} - \ln \left(\frac{l}{L}\right) + \frac{1}{2}\ln \left(\frac{\pi}{8}\right).
\ea
 In deriving this result we have taken ballot paths between the sites $0$ and $l$ ($h_l = 0$) and Dyck paths between the sites $l$ and 
$L$ ($h_L = 0$). One can compare the expression \rf{e1.2} with 
\rf{e31} 
(the same estimators for Dyck paths). 
One notices that the value of $\gamma_D=\gamma_S$ is the same but instead of having a constant like in 
 \rf{e28}, one has an asymmetric (under the interchange $l/L \leftrightarrow 1-l/L$) function of $l/L$. 
If $l/L$ approaches the value 1, one recovers the expression \rf{e29} as one 
should since one is far away from the "source" at the site $i = 0$. 
The limit $l/L \to 0$ is singular 
since for $L$ large there are no contact points for $l$ close to the 
origin.  We are going to find a similar situation in our next example.

 The raise and peel model with a wall (RPMW) is a simple extension \cite{APR} of the raise and peel model described in Sec.~5 . We consider this extension for the case $u = 1$ only. This is the case where one has conformal invariance. 

 The stochastic model is defined by the four rules given in Sec.5 to which we add the following rules: with a probability $P_i = 1/(L + a - 1)$ a tile from the gas hits the site $i$, ($i = 1,\ldots,L - 1$). 
The changes of the interface produced by the hits are the same as in the raise and peel model. With a probability $P_0 = a/(L + a - 1)$, a half-tile hits the site $0$. The boundary rate is equal to $a$ as opposed to the bulk rates which are equal to 1. The "source" rate $a$ stays as a free parameter in the problem. If the slope $s_0 = h_1 - h_0$  is equal to 1, the half-tile is adsorbed (see Fig.~\ref{fig8}). If $s_i = - 1$, the half-tile is reflected 
(see Fig.~10). 
\begin{figure}[t] 
\begin{center}
\begin{picture}(180,110)(0,-10)
{\green{
\qbezier[1](10,66)(10.5,66.5)(11,67)
\qbezier[2](10,64)(11,65)(12,66)
\qbezier[3](10,62)(11.5,63.5)(13,65)
\qbezier[4](10,60)(12,62)(14,64)
\qbezier[5](10,58)(12.5,60.5)(15,63)
\qbezier[6](10,56)(13,59)(16,62)
\qbezier[7](10,54)(13.5,57.5)(17,61)
\qbezier[8](10,52)(14,56)(18,60)
\qbezier[1](10,106)(10.5,106.5)(11,107)
\qbezier[2](10,104)(11,105)(12,106)
\qbezier[3](10,102)(11.5,103.5)(13,105)
\qbezier[4](10,100)(12,102)(14,104)
\qbezier[5](10,98)(12.5,100.5)(15,103)
\qbezier[6](10,96)(13,99)(16,102)
\qbezier[7](10,94)(13.5,97.5)(17,101)
\qbezier[8](10,92)(14,96)(18,100)}}
\qbezier[30](10,30)(25,45)(40,60)
\qbezier[40](10,10)(30,30)(50,50)
\qbezier[30](30,10)(45,25)(60,40)
\qbezier[20](50,10)(60,20)(70,30)
\qbezier[30](70,10)(85,25)(100,40)
\qbezier[20](90,10)(100,20)(110,30)
\qbezier[10](110,10)(115,15)(120,20)
\qbezier[10](130,10)(135,15)(140,20)
\qbezier[20](10,30)(20,20)(30,10)
\qbezier[40](10,50)(30,30)(50,10)
\qbezier[50](20,60)(45,35)(70,10)
\qbezier[20](70,30)(80,20)(90,10)
\qbezier[30](80,40)(95,25)(110,10)
\qbezier[10](120,20)(125,15)(130,10)
{\thicklines
\put(10,90){\line(1,1){10}}
\put(10,110){\line(1,-1){10}}
\put(10,90){\line(0,1){20}}
\put(10,88){\vector(0,-1){10}}
\put(10,70){\line(1,-1){10}}
\put(10,50){\line(0,1){20}}
\put(10,10){\line(1,0){140}}
\put(10,10){\line(0,1){40}}
\put(20,60){\line(1,1){10}}
\put(30,70){\line(1,-1){40}}
\put(70,30){\line(1,1){20}}
\put(90,50){\line(1,-1){30}}
\put(120,20){\line(1,1){10}}
\put(130,30){\line(1,-1){20}}
}
\put(10,50){\line(1,1){10}}
\qbezier[5](10,4)(10,7)(10,10)
\qbezier[5](20,4)(20,7)(20,10)
\qbezier[5](70,4)(70,7)(70,10)
\qbezier[5](120,4)(120,7)(120,10)
\qbezier[6](150,4)(150,7)(150,10)
\put(8,-3){$\scriptscriptstyle 0$}
\put(19,-3){$\scriptscriptstyle 1$}
\put(68,-3){$\scriptscriptstyle 6$}
\put(117,-3){$\scriptscriptstyle 11$}
\put(136,-3){$\scriptscriptstyle L=14$}
\end{picture}
\begin{picture}(180,110)(0,-10)
{\green{
\qbezier[1](10,46)(10.5,46.5)(11,47)
\qbezier[2](10,44)(11,45)(12,46)
\qbezier[3](10,42)(11.5,43.5)(13,45)
\qbezier[4](10,40)(12,42)(14,44)
\qbezier[5](10,38)(12.5,40.5)(15,43)
\qbezier[6](10,36)(13,39)(16,42)
\qbezier[7](10,34)(13.5,37.5)(17,41)
\qbezier[8](10,32)(14,36)(18,40)
\qbezier[8](12,50)(16,54)(20,58)
\qbezier[8](13,49)(17,53)(21,57)
\qbezier[8](14,48)(18,52)(22,56)
\qbezier[8](15,47)(19,51)(23,55)
\qbezier[8](16,46)(20,50)(24,54)
\qbezier[8](17,45)(21,49)(25,53)
\qbezier[8](18,44)(22,48)(26,52)
\qbezier[8](19,43)(23,47)(27,51)
\qbezier[8](20,42)(24,46)(28,50)
\multiput(0,0)(10,-10){3}{
\qbezier[8](22,60)(26,64)(30,68)
\qbezier[8](23,59)(27,63)(31,67)
\qbezier[8](24,58)(28,62)(32,66)
\qbezier[8](25,57)(29,61)(33,65)
\qbezier[8](26,56)(30,60)(34,64)
\qbezier[8](27,55)(31,59)(35,63)
\qbezier[8](28,54)(32,58)(36,62)
\qbezier[8](29,53)(33,57)(37,61)
\qbezier[8](30,52)(34,56)(38,60)}
\qbezier[8](52,75)(56,79)(60,83)
\qbezier[8](53,74)(57,78)(61,82)
\qbezier[8](54,73)(58,77)(62,81)
\qbezier[8](55,72)(59,76)(63,80)
\qbezier[8](56,71)(60,75)(64,79)
\qbezier[8](57,70)(61,74)(65,78)
\qbezier[8](58,69)(62,73)(66,77)
\qbezier[8](59,68)(63,72)(67,76)
\qbezier[8](60,67)(64,71)(68,75)
}}
\qbezier[30](10,30)(25,45)(40,60)
\qbezier[40](10,10)(30,30)(50,50)
\qbezier[30](30,10)(45,25)(60,40)
\qbezier[20](50,10)(60,20)(70,30)
\qbezier[30](70,10)(85,25)(100,40)
\qbezier[20](90,10)(100,20)(110,30)
\qbezier[10](110,10)(115,15)(120,20)
\qbezier[10](130,10)(135,15)(140,20)
\qbezier[20](10,30)(20,20)(30,10)
\qbezier[40](10,50)(30,30)(50,10)
\qbezier[50](20,60)(45,35)(70,10)
\qbezier[20](70,30)(80,20)(90,10)
\qbezier[30](80,40)(95,25)(110,10)
\qbezier[10](120,20)(125,15)(130,10)
{\thicklines
\put(50,75){\line(1,1){10}}
\put(50,75){\line(1,-1){10}}
\put(60,65){\line(1,1){10}}
\put(60,85){\line(1,-1){10}}
\put(60,63){\vector(0,-1){10}}
\put(10,10){\line(1,0){140}}
\put(10,10){\line(0,1){20}}
\put(60,40){\line(1,-1){10}}
\put(70,30){\line(1,1){20}}
\put(90,50){\line(1,-1){30}}
\put(120,20){\line(1,1){10}}
\put(130,30){\line(1,-1){20}}
\put(10,30){\line(1,1){20}}
\put(30,50){\line(1,-1){20}}
\put(50,30){\line(1,1){10}}
}
\put(10,30){\line(0,1){20}}
\put(10,50){\line(1,1){20}}
\put(30,70){\line(1,-1){30}}
\qbezier[5](10,4)(10,7)(10,10)
\qbezier[5](20,4)(20,7)(20,10)
\qbezier[5](70,4)(70,7)(70,10)
\qbezier[5](120,4)(120,7)(120,10)
\qbezier[6](150,4)(150,7)(150,10)
\put(8,-3){$\scriptscriptstyle 0$}
\put(19,-3){$\scriptscriptstyle 1$}
\put(68,-3){$\scriptscriptstyle 6$}
\put(117,-3){$\scriptscriptstyle 11$}
\put(136,-3){$\scriptscriptstyle L=14$}
\end{picture}
\parbox{15cm}{
\caption{\small  The adsorption of a half-tile at the first site
and desorption of a layer touching the boundary
for the
ballot path shown in Fig.\,9.}
}
\end{center}
\label{fig8}
\end{figure}
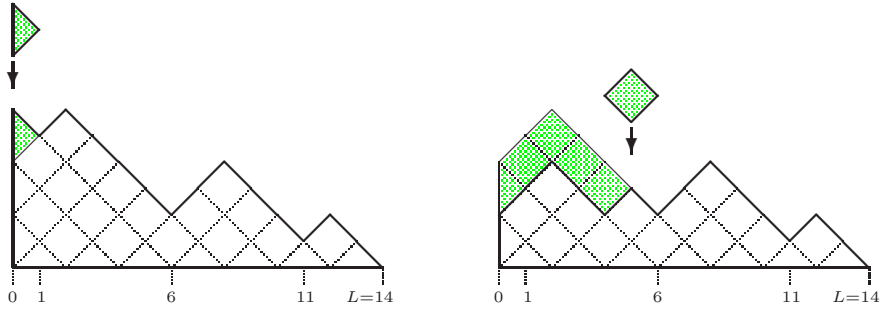
\begin{figure}[t]
\begin{center}
\begin{picture}(250,100)
\put(0,0){\epsfxsize=250pt\epsfbox{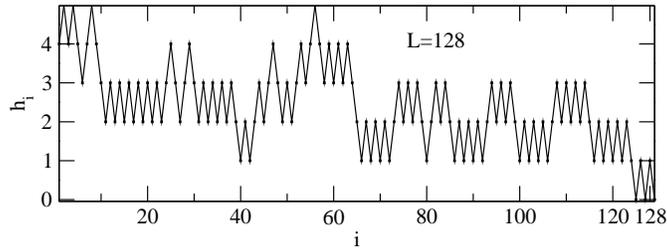}}
\end{picture}
\caption{\small Typical configuration in the stationary state for the RPMW with
a boundary rate $a = 1$. The system has a size $L = 128$. $h_i$ is the height
at site $i$. The contact points are concentrated at the right end.}
\label{fig9}
\end{center}
\end{figure}

 The properties of the stationary states of the raise and peel model with a wall have been described in detail in \cite{APR}. 
We are interested here in one result only. It was found  that in the finite-size scaling limit, for any boundary rate $a$, the density of contact points
has the expression: 
\be \label{e1.3}
\rho(l,L) \sim 0.8757 \left( \frac{1-\cos(\frac{\pi l}{L})}{L\sin(\frac{\pi l}
{L})}\right)^{1/3}.
\ee
From which we derive the following expression for the density of contact points estimator:
\be \label{e1.4}
D(l,L) \sim \frac{1}{3}\ln \tilde{L}_C + 
\ln\left[0.8757\left(1-\cos(\frac{\pi l }{L})\right)^{1/3}\right].
\ee
 One notices that for $l/L$ close to 1, one is back at the result 
\rf{e43} obtained in the raise and peel model 
and similar to Eq.~\rf{e1.2}, 
the limit $l/L \to 0$ is singular. We didn't look at other estimators.

\section{ Shared information in the asymmetric exclusion process with periodic boundary conditions.}

 To illustrate how the estimators defined in Section 2 can also be used for configurations which are not Dyck paths, we give a simple example: 
the asymmetric exclusion process on a ring.

 We do not describe here the very well known asymmetric exclusion process (see \cite{ASEP} for a review). We consider directly the stationary states.¬ For a ring with $L$ sites and $N$ particles, in the stationary state all
\be \label{e1.5}
Z(N,L) = \frac{N!}{N!(L-N)!}
\ee
configurations have the same probability. We are interested in the shared information between a segment with $l$ sites and the remaining segment with 
$L-l$ sites on the ring. The average local density is $r = N/L$. The shared information between the two segments comes from the fact that if we measure $Q$ particles in the segment $l$, one has to have N-Q particles in the segment $L - l$. 
 
 The {\it mutual information},  equal to the {\it interdependency} (see 
\rf{e15}), has  the expression:
\be \label{e1.6}
I(l,L) = H_q (l,L) = -\sum_Q P_l(Q,L) \ln P_l(Q,L), 
\ee
where 
\be \label{e1.7}
P_l(Q,L) = Z(Q,l)Z(N-Q,L-l)/Z(N,L).
\ee  
Denoting $Q = rl + q$ and taking  $|q|/L << 1$, in the finite-size 
scaling limit, $P_l(q,L)$ is given by a Gauss probability distribution:
\be \label{e1.8}
P_l(q,L) \sim \frac{1}{\sigma\sqrt{2\pi}} \exp(-\frac{q^2}{2\sigma^2}),
\ee
where 
\be \label{e1.9}
\sigma= \sqrt{r(1-r)\tilde{L}_{RW}},
\ee
and  $\tilde{L}_{RW}$ is given by \rf{e27b}. 
From \rf{e1.6} and \rf{e1.8} one obtains.
\ba \label{e1.10}
&&I(l,L) = H_q(l,L) 
\sim\frac{1}{2}\ln\tilde{L}_{RW} +\frac{1}{2} \ln(2\pi r(1-r)) +\frac{1}{2}.
\ea
 The equivalent of the {\it valence bond entanglement entropy} is:
\be \label{e1.11}
q(l,L) = \sum_q |q| P_q(l,L) \sim 2 \sigma =
2\sqrt{r(1-r)} \tilde{L}_{RW}^{1/2}.
\ee
 Finally, the equivalent of the {\it density of contact points} estimator, equal to the {\it separation Shannon entropy} is:
\ba \label{e1.12}
&&D(l,L) = S(l,L) = -\ln P_l(0,L) 
\sim \frac{1}{2}\ln \tilde{L}_{RW} + 
\frac{1}{2}\ln (2\pi r(1-r)). 
\ea

 We notice that the expressions of the estimators are very similar to those obtained in the case of the model of the polymer adsorption in the domain $0 < K < 2$. One difference is that in the present case, four of the estimators have the same leading expression and that the common coefficient 
$\gamma_E$
  is smaller (one has to take half the value  obtained in the present case since we have periodic boundary conditions). 
Moreover, half of the coefficient in front of $L_{RW}^{1/2}$ in 
\rf{e1.12} 
is smaller than the coefficient
of $L_{RW}^{1/2}$ in \rf{e30}.
This implies that the shared information is smaller than in the polymer adsorption model. 

Another difference is that the coefficient in front of $\tilde{L}_{RW}^{1/2}$ 
in \rf{e1.11} depends on the density of particles $r$.  The $\gamma _E$'s 
coefficients (equal 
to 1/2) are not. The $r$-dependence being hidden in the constant term.

\section{ Conclusions.}

 Our aim was to see if for stationary states of one-dimensional Markov 
processes with many degrees of freedom, one can define estimators 
of shared information between two subsystems. 
We wanted to make a connection between information theory and 
non-equilibrium statistical physics paralleling the 
connection  between quantum information theory and extended quantum systems. We found several useful estimators. 
The Markov process gives the rules how to perform Monte Carlo simulations. 
These simulations on large system sizes were used to compute some 
estimators. 
This method can not be used in the quantum case.

 Although we consider classical systems, it turns out that the 
estimators have properties similar to their counterparts in the 
quantum mechanical case.

 The estimators are defined in Sec.~2. We have studied their properties 
in several models described in Sections 3, 4, 6 and Appendix B. 
If the subsystems have lengths $l$ and $L-l$, all the estimators  
have the following behavior in the finite-size scaling limit 
($1 << l$,$L$, $l/L$ fixed):

a) They vanish if the subsystems are separated (for any $l$ and $L$).

b) If the correlation length is finite, they stay finite for large values of
 $L$ (area law).

c) If the system is critical, for each universality class one can 
define a characteristic length
\be \nonumber
\tilde{L} =  L f(x),\quad  f(x) = f(1-x), \quad f(x) \sim x \quad (x <<1), 
\ee
where $x = l/L$ and $f(x)$ depends on the universality class.

d) At criticality one obtains corrections to the area law. 
They all depend on $\tilde{L}$ only and can increase logarithmically or 
as  a power of $\tilde{L}$.

e) If one has logarithmic corrections, the estimators 
$E(l,L)$ have the 
following expression:
\be \label{e72}
E(l,L) \sim \gamma_E \ln \tilde{L} + C_E,
\ee  
where $\gamma_E$ depends on the estimator but is universal, 
$C_E$ is not universal.

f) If one has power-law corrections, the estimators have the following 
expression:
\be \nonumber
E(l,L) \sim \delta_E (\tilde{L})^m + D_E,     
\ee
where the exponent $m < 1$ and the constant $\delta_E$ are universal 
and $D_E$ is not universal.

g) If one compares the coefficients $\gamma_E$ and $\delta_E$  belonging to two 
universality classes, 
they are all larger (smaller).

 In our study we have found only one estimator (the valence bond 
entanglement entropy 
\rf{e9}) for which one gets power-law corrections with $m = 1/2$. 
This
 was seen in the models discussed in Sec.~3
and 6 but not in the model of Sec.~5. 
All the other estimators give logarithmic corrections. 
We don't yet have  an explanation of this observation.

 We do have some understanding of the physical ingredients 
which enter the expression of the estimators.

 The simplest case is when we have conformal invariance ($u = 1$ in the raise
 and peel model of
Sec. 4). The expression of $\tilde{L} = L_C$ 
 \rf{e53}
 is the same as in the quantum case
 and is well understood \cite{CAR}. The value of $\gamma_h$ (see Table 1) 
can be inferred from the calculation of Jacobsen and Saleur 
\cite{JLJ} for the periodic case. 
$\gamma_D$ is related to the critical exponent of the local density 
of contact points which is known exactly. 
The values of $\gamma_I$, $\gamma_H$ and $\gamma_S$ have
still to be derived analytically. Subleading contributions to 
 \rf{e72} 
observed if one considers even or odd sites are discussed in Sec.~5. In
the case of the R\'enyi interdependencies, a new, unexpected phenomenon is
seen: the differences between the R\'enyi interdependencies for odd and even
sites oscillate.

 We do not have  a clue how to derive the values of $\gamma_E$'s or the 
functions $\tilde{L}_u$ for the whole domain $u > 1$ of the 
raise and peel model.

 It looks as if  the models of Sec. 3, 6 and Appendix B have one 
common $\tilde{L}$ with $f(x) = x(1-x)$. 
These models share a common feature: the heights respectively the 
hopping particles have Gauss fluctuations.

 We have also studied simple examples of left-right 
asymmetric partitions and found the finite-size scaling 
functions in these cases.

 The estimators defined here for bipartitions can be extended in an 
obvious way to multipartitions. 
The simplest case is the separation (contact) points of tripartitions,  
which is related to a not connected two-point correlation function. 
We did not touch this topic in this
paper.

 We hope to see the application of the methods introduced here to other 
stochastic processes. Probably new aspects of the shared 
information theory will
 unravel.

\section*{Acknowledgments}
G. Sierra has participated in the early stages of our investigations and we
are grateful for his contribution. 
We would like to thank P. Pyatov for providing us with  his calculations 
 of the mutual information and the separation Shannon entropy for the  raise and 
peel model at $u=1$, with lattices sizes 
up to $L=12$, and also for related discussions. We are also grateful to A. L Owczarek for 
 making  available to us his manuscript and to P. Calabrese for discussions. 
 This work  was
partially supported by FAPESP and CNPq (Brazilian Agencies).

\appendix
\section{ Polymer adsorption for seven sites}

 In this appendix we derive the Hamiltonian and the expression of the 
stationary state for the model described in Section 4  in the case $L = 6$.
\begin{figure}[t]
\begin{center}
\begin{picture}(250,100)
\put(0,0){\epsfxsize=250pt\epsfbox{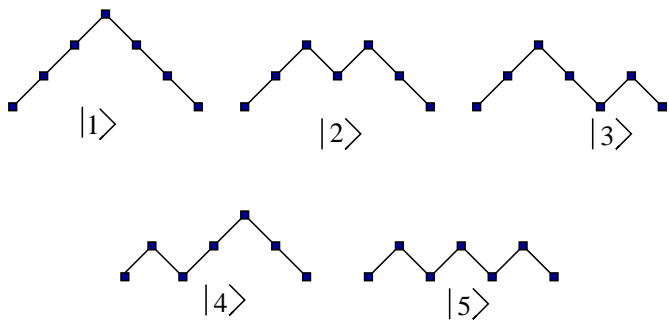}}
\end{picture}
\caption{\small The five Dyck path configurations for $L=6$}
\label{figa1}
\end{center}
\end{figure}

 There are 5 Dyck paths configurations in this case shown in Fig.~\ref{figa1}. The time evolution of the probabilities $P_a(t)$ to find the system in 
the state $\ket{a}$ ($a =1,2,\ldots,5$) at the time $t$ is given by the 
$5 \times 5$ matrix $H_{a,b}$. Its non-diagonal matrix elements  
are given by the rates $b \rightarrow  a$ with the sign changed 
($H_{a,b} \leq 0$) and the diagonal elements are fixed from the conditions: $\sum_a H_{a,b} = 0$.

 Denoting 
\be \label{a1}
\ket{P(t)} =\sum_{a =1}^5 P_a(t)\ket{a}, \quad P_a = \lim_{t \to \infty} 
 P_a(t), \quad \ket{0} = \sum_{a=1}^5 P_a\ket{a}
\ee
we get
\be \label{a2}
d/dt \ket{P(t)} = -H\ket{P(t)}, \quad H\ket{0} = 0.
\ee
The Hamiltonian with the transitions rates fixed by the rules given in 
Section 4 is:
\ba \label{a3}
&& H = 
\left( \begin{array}{c|rrrrr}
 & \ket{1} & \ket{2} & \ket{3} & \ket{4} &\ket{5} \\ \hline
\bra{1} & 1  & -1 & 0    & 0     & 0  \\
\bra{2} & -1 & 3  & -$u$ &  -$u$ & 0 \\ 
\bra{3} & 0  & -1 &$1+u$ & 0     & -$u$ \\
\bra{4} & 0  & -1 & 0    & $1+u$ & -$u$ \\
\bra{5} & 0  &  0 & -1   &  -1   & 2$u$  
\end{array} \right) .
\ea
 The stationary state follows:
\be \label{a4}
\ket{0} = \ket{1}+\ket{2}+u^{-1}(\ket{3}+\ket{4}) + u^{-2}\ket{5}.
\ee

\section{  
 Restricted and unrestricted Motzkin paths.}

 A Motzkin path is defined as follows. Consider the Cartesian plane 
$Z \times Z$ with the sites on the $x$-axis and the heights on the 
$y$-axis. The lattice paths start at $(x,y)=(0,0)$  and one can use steps 
$(U',L',D')$, 
where $U' =(1,1)$ is an up-step, $L' = (1,0)$ is a level-step and 
$D' = (1,-1)$ is a down-step. The steps end at $(x,y)=(L,0)$ without going below the $x$-axis. 
There are 
\be \label{b1} 
Z_M (L) = \sum_{k = 0}^{L/2} \frac{L!}{k!(k+1)!(L-2k)!} 
\ee
configurations of this kind. For large values of $L$ 
one obtains \rf{e38}.

 If one takes the $Z_M (L)$ Motzkin paths with equal probabilities, 
they describe a random walker who starts and returns to the origin  
after $L$ steps moving with a  diffusion constant $D$   differing
 from 
the 
 Dyck paths cases. This explains \rf{e39}.
The leading finite-size behavior of the estimators is independent of $D$.

 The restricted Motzkin paths are defined for $L = 3n$ and are 
subjected to the constraint that for any $0 < x < L$ one has 
$N_{U'} (x) \geq N_{L'} (x) \geq N_{D'} (x)$. This means that after $x$ steps, 
the number of up-steps $N_{U'}$  has to be larger or equal than the number of 
level-steps $N_{L'}$ which are larger or equal the number of down-steps 
$N_{D'}$. 

 The number of restricted Motzkin paths is equal to \cite{FZJ}:
\be \label{b2}
Z_{RM} (L) = \frac{2L!}{n!(n+1)!(n+2)!}.
\ee
 There are obviously fewer restricted Motzkin paths than unrestricted ones. For example if $L = 3$ one has only one restricted path as opposed to 
four unrestricted ones. The number of restricted Motzkin paths is 
equal to the number of independent $sl(3)$ singlet configurations 
one gets in a one-dimensional lattice with $L+1$ sites, if one puts on 
each site the 3-dimensional fundamental representation of $sl(3)$.

 Due to the fact that  restricted Motzkin configurations are 
more constrained than the Dyck paths, one expects in the case of a 
bipartition, larger  shared information between the two subsystems.
 This is indeed the case.
 Using the \rf{e15} 
 with $Z_1(L)$ replaced by $Z_{RM}$ one obtains the following 
finite-size scaling expression for the separation entropy 
(equal, in this case, to the density of separation points estimator):
\be \label{b3}
S(p,n) = 4 \ln [p(1-p/n)] + \ln(\pi/\sqrt{3}),
\ee
where we have taken $L = 3n$ and $l = 3p$ the number of sites in the 
bipartition. Comparing \rf{b3} to 
\rf{e31} we notice that the finite-size scaling function is 
unchanged but the value of $\gamma_S$ jumped from 3/2 for Dyck paths to 4 in the case of restricted Motzkin paths.

\end{document}